\documentclass[twocolumn,showpacs,preprintnumbers,amsmath,amssymb]{revtex4}
\usepackage{graphicx}% Include figure files
\usepackage{dcolumn}% Align table columns on decimal point
\usepackage{bm}% bold math

\begin{document}

%\preprint{to be submitted to PRL}

\title{Phase coherent transport in Kondo/superconducting hybrid structures}% Force line breaks with \\

\author{Jonghwa Eom}
\email{eom@sejong.ac.kr} \affiliation{Department of Physics,
Sejong University, Seoul 143-747, Korea}

\author{Yun-Sok Shin, Hu-Jong Lee, and Wang-Hyun Park}
\affiliation{Department of Physics, Pohang University of Science
and Technology, Pohang 790-784, Korea}

\author{Taegon Kim and Jonghan Song}
\affiliation{Advanced Analysis Center, Korea Institute of Science
and Technology, Seoul 130-650, Korea}

\date{\today}% It is always \today, today,

\begin{abstract}
We present measurements of the transport properties of hybrid
structures consisting of a Kondo AuFe film and a superconducting
Al film. The temperature dependence of the resistance indicates
the existence of the superconducting proximity effect in the Kondo
AuFe wires over the range of $\sim0.5$ $\mu$m. Electronic phase
coherence in the Kondo AuFe wires has been confirmed by observing
the Aharanov-Bohm effect in the magnetoresistance of the loop
structure. The amplitude of the magnetoresistance oscillations
shows a reentrant behavior with a maximum at $\sim$ 870 mK, which
results from an interplay between the Kondo effect and the
superconducting proximity effect.
\end{abstract}

\maketitle

Phase coherence has been a prime subject of interest in mesoscopic
systems. The phenomena of electron phase coherence have been
widely investigated in low-dimensional structures such as quantum
dots~\cite{Yacoby95}, carbon nanotubes~\cite{Kong01}, and metal
films~\cite{Mohanty97}. It is well known that phase coherence is
affected by the coupling of electrons to an environment. In a
system containing magnetic impurities, the coupling between the
electron spin and the magnetic impurity spin provides a very
efficient source of decoherence. Magnetic impurities of
fluctuating spins randomize the phase of electrons during the
scattering process. A small amount of magnetic impurities can
greatly shorten the phase coherence time, $\tau_{\phi}$. In a
recent experiment~\cite{Schopfer03} of a mesoscopic
dilute-magnetic-impurity system, however, $\tau_{\phi}$ was seen
to be as long as a few nsec at low temperatures. In addition,
phase coherence was reported to be preserved in a semiconductor
quantum dot in the Kondo regime~\cite{Wiel00}. Thus, one may
expect an observation of the phase coherent transport in a
diffusive metallic Kondo wire.

The Kondo effect originates from spin-flip scattering mediated by
the exchange coupling between a conduction-electron spin and a
localized impurity spin, which leads to a logarithmic increase of
the low-temperature resistance. For a dilute-magnetic-impurity
system in contact with a superconductor, the Kondo effect can
compete with the superconducting proximity effect. The microscopic
mechanism of the superconducting proximity effect is the Andreev
reflection: an incident electron reflects as a hole at the normal
metal/superconductor interface, simultaneously producing a Cooper
pair which propagates into the superconductor. Possible
coexistence of both effects was studied for systems where the
Kondo temperature $T_{K}$ was comparable to the superconducting
gap energy~\cite{Buitelaar02,Avishai03}.

\begin{center}
\begin{figure}
\includegraphics[width=8.2cm]{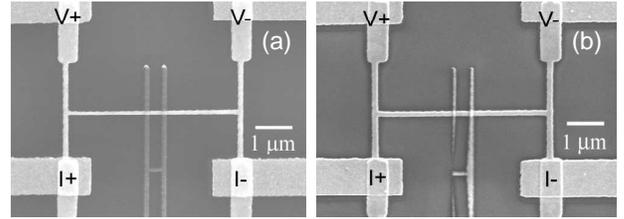}
\caption{Sample geometry. (a) Scanning electron micrograph of
sample $A$. A horse-shoe-type Al wire makes a hybrid loop at the
center of a AuFe wire. The AuFe film appears brighter than the Al
film. (b) Scanning electron micrograph of sample $B$. Since the Al
wire was broken, no hybrid loop is present.}
\end{figure}
\end{center}

In this paper we report the observation of phase coherent
transport through a mesoscopic Kondo system, to which
superconducting wires are connected with high transparency. The
Kondo system is a AuFe wire with a Fe concentration of 26 ppm, and
Al film is used for a superconductor. The number of Fe ions in the
1-$\mu$m-long AuFe wire of this experiment is estimated to be
approximately 1800. The Kondo effect is confirmed by logarithmic
temperature dependence of the resistivity at low temperatures. In
addition, the superconducting proximity effect has been observed
in the AuFe wire when the temperature, $T$, is lowered below the
superconducting transition temperature, $T_{c}$, of the Al films.
The resistance starts to drop as the samples are cooled through
$T_{c}$ and continuously decreases until $T$ reaches $\sim$ 0.3 K.
Phase coherent transport has been confirmed by the
magnetoresistance oscillations of the hybrid loop consisting of a
AuFe wire and an Al wire. More interestingly, it has been found
that the amplitude of the magnetoresistance oscillations shows a
strong temperature dependence. The oscillation amplitude initially
increases as $T$ is decreased, showing a maximum around $\sim$ 870
mK, and then decreases rapidly as $T$ is further lowered. The
non-monotonic temperature dependence of the magnetoresistance
oscillations is attributed to an interplay between the Kondo
effect and the superconductivity.

The samples in this experiment were patterned by using the
multilevel electron-beam lithography and lift-off process. In the
first lithography step, a 500-\AA-thick film was deposited by
thermal evaporation of 99.999\%-pure Au. After lift-off, the pure
Au film was implanted with Fe ions to a concentration of $\sim$ 26
ppm~\cite{FeImplant}, which was estimated subsequently from the
slope of the temperature-dependent resistivity of a control AuFe
wire~\cite{FeConcent}. In the second lithography step, a
1200-\AA-thick Al film was deposited to make hybrid structures in
the middle of the sample wires, the length of which was $\sim$ 4.7
$\mu$m. The area of the interface between the AuFe film and the Al
film was approximately 0.12 $\times$ 0.14 $\mu$m$^{2}$. The
interface resistance was estimated to be $\sim$ 0.1 $\Omega$.
Figure~1 shows the scanning electron micrographs of two hybrid
structured samples. Sample $A$ is shown in Fig.~1(a), and sample
$B$ is shown in Fig.~1(b). In sample $A$, a horse-shoe-type Al
wire was located at the center of the AuFe wire, making a
rectangular hybrid loop composed of a superconductor and a Kondo
wire. However, in sample $B$, one arm of the rectangular hybrid
loop was not lithographically completed, so the Al contacts made
two separate interfaces with the AuFe wire.

The samples were measured in a dilution refrigerator using
standard lock-in techniques with a four-terminal ac resistance
bridge. The four-terminal measurement configurations for the two
samples are described in Fig.~1(a) and (b), respectively. The
resistivity of the control AuFe wire at 4.2 K was 1.37
$\mu\Omega$cm, and the thermal length of the AuFe wire, $L_{T} =
\sqrt{\hbar D /k_{B}T}$, was $\sim$ 0.47/$\sqrt{T[K]}$ $\mu$m at a
temperature $T$. Here, $D$ is the diffusion
constant.

\begin{center}
\begin{figure}
\includegraphics[width=7.4cm]{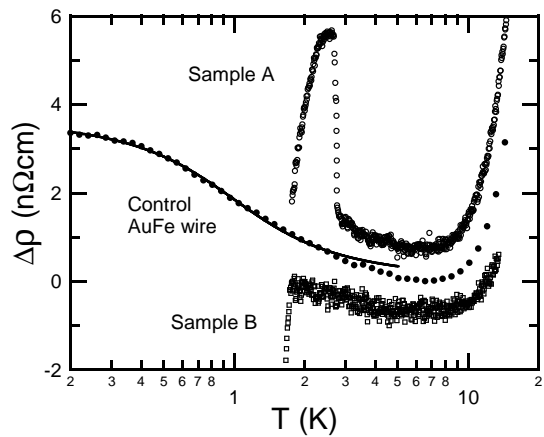}
\caption{Temperature dependence of the resistivity of sample $A$
(open circle), a meander-type control AuFe wire (closed circle),
and sample $B$ (open square) in zero magnetic field. The minima of
resistivity occur at $T \approx$ 7.0 K for all samples, and the
resistivities at 7.0 K are 1.40 $\mu\Omega$cm, 1.37 $\mu\Omega$cm,
and 1.20 $\mu\Omega$cm for sample $A$, the AuFe control wire, and
sample $B$, respectively. The solid line represents a fit to the
Hamann function with $T_{K}$= 0.99 K.}
\end{figure}
\end{center}

Figure~2 shows the zero-magnetic-field temperature dependent
resistivity of sample $A$, $B$, and the control AuFe wire. The
control AuFe wire was co-fabricated with samples on the same
substrate by using the simultaneous lithographical processes and
Fe implantation. The control AuFe wire has the dimensions of
374-$\mu$m length and 0.14-$\mu$m width. At temperatures above
$\sim$ 7.0 K, where phonon contribution dominates the resistivity,
the resistivity of the AuFe wire decreases as $T$ is lowered. For
$T$ below $\sim$ 7.0 K, the resistivity $\rho(T)$ of the control
AuFe wire shows the Kondo effect. For the resistivity of dilute
magnetic alloys, the lowest-order calculation in the second Born
approximation yields a term linear in $\log T$. When the Kondo
temperature $T_{K}$ is compared to the experimental temperatures,
the purely logarithmic dependence is slightly modified.
Considering the nature of the spin correlations over the
temperature range above and below $T_{K}$, Hamann derived a
specific functional form for the Kondo contribution to
$\rho(T)$~\cite{Hamann67}:
\begin{equation}\label{HamannEq}
\Delta \rho(T) \sim \frac{1}{2} \rho_{0} \left(1 \pm \left[
1+\frac{S(S+1)\pi^2}{[\ln (T/T_{K})]^2} \right]^{- \frac{1}{2}}
\right),
\end{equation}
where the positive sign is for $T < T_{K}$, the negative sign is
for $T > T_{K}$, and $\rho_{0} = 4\pi c \hbar / ze^{2}k_{F}$ with
$c$ being the concentration, $z$ the number of conduction
electrons per atom, and $k_{F}$ the Fermi wave vector. And, $S$ is
the effective spin of the magnetic impurity in a host metal. The
Hamann expression in Eq.~(1) readily explains the parabolic
dependence of $\rho(T)$ for $T \ll T_{K}$. By fitting to Eq.~(1),
one can determine $T_{K}$ of a dilute magnetic alloy. The solid
line in Fig.~2 represents a fit to the Hamann function with $S$ =
0.12 and $T_{K}$ = 0.99 K. The magnitude of $T_{K}$ is comparable
to the known values from the previous studies on the
dilute-magnetic-impurity system of AuFe~\cite{Daybell73,Loram70}.

The resistivity for samples $A$ and $B$ also increases as $T$ is
lowered below $\sim$ 7.0 K, yielding a similar feature of
resistivity minimum as in the control AuFe wire. The temperature
dependence of $\rho(T)$ ensures the presence of the Kondo effect
in samples $A$ and $B$. In addition, however, a sharp drop of
$\rho(T)$ occurs for both samples as $T$ is lowered below $T_{c}$
of the Al film, which is approximately 1.6 K in this experiment.
For sample $A$, an anomalous peak is observed at temperatures
above $T_{c}$. Similar resistance anomalies have been reported
near the superconducting transition of mesoscopic Al
samples~\cite{Santhanam91,Strunk98,Kim94}. Explanations in terms
of nonequilibrium charge imbalance around phase-slip
centers~\cite{Arutyunov96} and pinching of the conducting path
near the nodes of the voltage leads~\cite{Kim94} were proposed. In
sample $B$ the current path does not include the Al wire
explicitly, so the resistance anomaly is not observed.

Since Giroud $et$ $al.$ reported an anomalous temperature
dependence in the resistivity of mesoscopic Co/Al hybrid
structures~\cite{Giroud98}, existence of superconducting proximity
effect in magnetic systems has been a subject of continuing
experimental and theoretical interest~\cite{Petrashov99,Belzig00}.
In a highly transparent interface (high conductance), the
superconducting pair correlations can penetrate into a magnetic
metal and give rise to the proximity effect. For ferromagnetic
metals, the proximity effect extends to a distance determined by
the exchange field energy, $\xi_{ex}$ = $\sqrt{\hbar D /
k_{B}T_{Curie}}$, where $T_{Curie}$ is the Curie temperature of a
ferromagnetic metal. This length scale can be as long as $\sim$ 10
nm for weak ferromagnetic metals such as Cu-Ni
alloys~\cite{Ryazanov01} or Pd-Ni alloys~\cite{Kontos01}.

\begin{center}
\begin{figure}
\includegraphics[width=7.4cm]{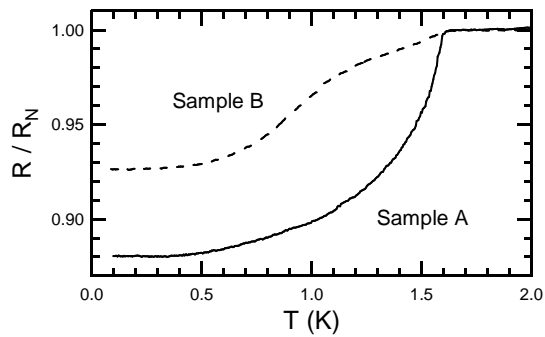}
\caption{The normalized resistance $R/R_{N}$ for samples $A$
(solid line) and $B$ (dashed line) as a function of temperature.
$R_{N}$= 11.5 $\Omega$ and $R_{N}$= 8.7 $\Omega$ for sample $A$
and $B$, respectively. The anomalous peak in the resistance of
sample $A$ still exists above $T_{c}$, but it is not clearly seen
in the scale of this figure.}
\end{figure}
\end{center}

For a dilute-magnetic-impurity system, the superconducting pair
correlations are likely to penetrate into a normal metal even
longer than for a weak ferromagnetic metal. The proximity effect,
which is sensitive to $T$, would cause a continuous change of
resistance at temperatures below $T_{c}$. Such a temperature
dependence of the resistance has been found for both samples $A$
and $B$, as shown in Fig.~3. The normalized resistance in zero
magnetic field begins to decrease rapidly as $T$ is lowered below
$T_{c}$ of the Al film. Since a part of the AuFe wire in sample
$A$ is shorted by the superconducting Al arm, a resistance drop of
sample $A$ is larger than that of sample $B$. Following a slow
initial decrease, the resistance of sample $B$ decreases rapidly
below $T \sim$ 1.1 K, which may be associated with Josephson
coupling between two separate Al contacts in sample $B$.

Since an injected electron below the supercondcuting gap energy is
Andreev-reflected with phase memory of the superconducting
condensate, the resistance of a normal metal wire between two
normal metal/superconductor (N/S) interfaces can be modulated by
the macroscopic phase of the superconducting condensate. An easy
way to manipulate the phase is achieved by applying a magnetic
field through the loop geometry. However, a prerequisite for the
resistance modulation is phase coherence in the normal metal
between the two N/S interfaces. Therefore, the magnetoresistance
oscillations in the AuFe/Al hybrid loops provide direct evidence
of phase coherent transport in the Kondo AuFe wire. The strong
Aharonov-Bohm effect shown in Fig.~4(a) in the magnetoresistance
of the hybrid loop (sample $A$) presents such evidence. The
oscillation period of $\Delta B \approx$ 25.4 G corresponds to the
superconducting flux quantum, $\Phi_{0}$ ($= h /2e$), divided by
the loop area, which is equal to the area enclosed by the center
of the hybrid loop ($\approx 0.80 \pm 0.01 \mu$m$^2$).

\begin{center}
\begin{figure}[t]
\includegraphics[width=7.4cm]{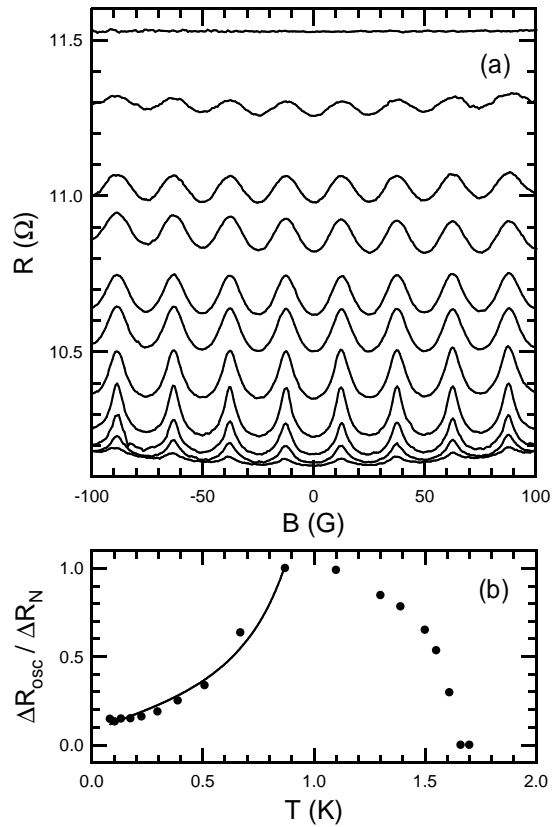}
\caption{(a) The magnetoresistance for sample $A$ at $T$= 83 mK,
508 mK, 670 mK, 870 mK, 1.1 K, 1.3 K, 1.39 K, 1.5 K, 1.55 K, 1.61
K, 1.66 K from bottom to top. (b) The normalized amplitude of the
magnetoresistance oscillations for sample $A$ as a function of
temperature. The normalization constant $\Delta R_{N}$ is 0.16
$\Omega$. The solid line represents a fit to the form $(a +
b\log(T/T_{K}))^{-1}$ where $T_{K}$ = 0.99 K, with $a$ = 0.57 and
$b$ = -3.21.}
\end{figure}
\end{center}

Figure~4(b) shows the amplitude of the magnetoresistance
oscillations for sample $A$. The amplitude was determined by the
difference between $R$ ($B$=12.7 G) and $R$(0). The amplitude
shows a reentrant behavior with a maximum at $T_{m} \sim$ 870 mK.
The temperature dependence above $T_{m}$ is related to the
superconducting proximity effect, which becomes stronger as $T$ is
lowered. However, the temperature dependence of $\Delta R_{osc}$
qualitatively different from the data previously reported for the
loops consisting of nonmagnetic metal films and superconducting
films~\cite{Courtois96,Chien99}. In the previous reports the
oscillation amplitudes were seen to decay as either a power law or
an exponential form in temperature. The peculiar temperature
dependence above $T_{m}$ in Fig.~4(b) is likely due to the
superconducting proximity effect in the presence of spin-flip
scattering of the conduction electrons on the magnetic impurities.

At lower temperatures the $s-d$ exchange coupling between a
conduction electron spin and a localized impurity spin makes the
spin-flip scattering even more enhanced, leading to the so-called
Kondo effect. As the spin-flip scattering increases, electron
phase coherence becomes weaker. Consequently the amplitude of the
magnetoresistance oscillations decreases as the temperature is
further lowered. Assuming that the scattering rate in a dilute
magnetic system is decomposed into a term for spin-flip scattering
and another for all other processes, one obtains a rough
description for the temperature dependence of the amplitude below
$T_{m}$ in Fig.~4(b). The scattering rate is given by
$\tau_{0}^{-1} + b\log (T/T_{K})$, where $\tau_{0}$ represents the
effective scattering time originating from all other processes
except for the spin-flip scattering, and $b$ is a negative
constant. Considering spin-flip scattering dominates the
low-temperature decoherence, $\tau_{0}$ is assumed to be
independent of temperature in our rough estimate. As the
scattering rate increases, phase coherence of the conduction
electron becomes weaker. The amplitude of the magnetoresistance
oscillations, $\Delta R_{osc}$, is expected to be inversely
proportional to the scattering rate: $\Delta R_{osc} \propto (a +
b\log (T/T_{K}))^{-1}$. The solid line in Fig.~4(b) represents a
fit to the above relation with $a$ = 0.57 and $b$ = -3.21, keeping
$T_{K}$ = 0.99 K for the AuFe wires in this experiment. The fit
describes $\Delta R_{osc}$($T$) reasonably well in the temperature
range below $T_{m}$.

Mesoscopic length scales which play an important role in this
experiment are the thermal length, $L_{T}$, and the phase
coherence length, $L_{\phi}$. $L_{T}$ of the AuFe wires is
estimated to be $\sim$ 0.47/$\sqrt{T[K]}$ $\mu$m. Characterized by
$L_{T}$, the range of the superconducting proximity effect already
extends over the entire AuFe arm ($\sim$ 0.5 $\mu$m long) of the
hybrid loop when the temperature reaches $T_{c}$ of the Al film.
However, the superconducting proximity effect is also bound by the
phase coherence length, $L_{\phi}$, which is affected by the
spin-flip scattering on the magnetic impurities. The Aharonov-Bohm
effect in the magnetoresistance is disturbed as $L_{\phi}$ becomes
shorter than the length of the normal arm of a hybrid loop.
Although a specific estimate of $L_{\phi}$ is not available in the
present work, the existence of magnetoresistance oscillations
indicates that $L_{\phi}$ in the Kondo AuFe wire below $T_{c}$ is
longer than the separation between the two N/S interfaces in
sample $A$. This is consistent with the previous estimate by
Schopfer \emph{et al.} of the phase coherence time in
quasi-one-dimensional AuFe wires~\cite{Schopfer03}. In future
studies on AuFe films with various Fe concentrations in close
proximity to a superconductor, we expect to elucidate the
characteristic interplay between $L_{T}$ and $L_{\phi}$ in the
superconducting proximity effect.

We thank T.-S. Kim and H.-W. Lee for helpful discussions. This
work was supported by the KOSEF through the electron Spin Science
Center at POSTECH.

%\newpage %Just because of unusual number of tables stacked at end

\begin{thebibliography}{}
\bibitem{FeImplant} The samples were implanted at dose of 1.3
$\times$ $10^{13}$ ions/cm$^{2}$ and energy of 80 keV.
\end{thebibliography}

\end{document}